\documentclass[pra,aps,twocolumn]{revtex4}

\usepackage{amssymb}
\usepackage{amsmath}
\usepackage{epsfig}

\begin{document}
\title{Rank three bipartite entangled states are distillable}
\author{Lin Chen}
\author{Yi-Xin Chen}
\affiliation{Zhejiang Insitute of Modern Physics, Zhejiang
University, Hangzhou 310027, China}

\begin{abstract}
We prove that the bipartite entangled state of rank three is
distillable. So there is no rank three bipartite bound entangled
state. By using this fact, We present some families of rank four
states that are distillable. We also analyze the relation between
the low rank state and the Werner state.
\end{abstract}
\maketitle

\section{Introduction}
In recent years, entanglement has been regarded as a quantum
resource for many novel tasks such as quantum computation, quantum
cryptography, quantum teleportation and so on \cite{Nielsen}.
These quantum-information tasks cannot be carried out by classical
resources and they rely on the entangled states. Although the
mixed entangled states are directly used in some
quantum-information tasks \cite{Murao}, most of them require the
pure entangled states of bipartite or multipartite system to be
the crucial elements. However in a lab, it turned out that the
pure entangled states always become mixed by the decoherence due
to the coupling with the environment. A central topic in quantum
information theory is thus how to extract pure entangled states
from mixed states \cite{Horodecki1}.

An entangled state $\rho$ is distillable if one can asymptotically
or explicitly extract some pure entangled state from infinitely
many copies of $\rho$ by using only local operations and classical
communication (LOCC). It has been proved that the entangled
2-qubit states are always distillable
\cite{Bennett1,Bennett2,Horodecki2}. Nevertheless there exist
bound entangled (BE) states which are not distillable under LOCC
\cite{Horodecki4}. Concretely, a bipartite entangled state $\rho$
in the Hilbert space $H_A\otimes H_B$ is BE if it has positive
partial transpose (PPT) with respect to system $A$ (or $B$),
namely $\rho^{T_A}(\mbox{or}\ \rho^{T_B})\geq0$. Such states are
called PPT BE states and usually it cannot be used for
quantum-information tasks under LOCC \cite{Murao,Eggeling}.

A more formidable challenge is that whether a bipartite state
$\rho_{AB}$ having non-positive partial transpose (NPT) with
respect to system $A$ (or $B$) is always distillable. This class
of states are always entangled due to the celebrated
Peres-Horodecki criterion \cite{Peres}. It was pointed out by
\cite{Horodecki5} that any NPT state can be converted into some
NPT Werner state under LOCC. Much efforts have been devoted to
distilling this kind of states and there has been a common belief
that NPT BE Werner states indeed exists
\cite{DiVincenzo,Vianna,Pankowski,Hiroshima,Bandyopadhyay,Kraus,Clarisse}.
In addition, it has been proved that the NPT states in $2\times N$
space are distillable \cite{Horodecki2,Kraus2} and the rank two
NPT states of bipartite systems are also distillable
\cite{Horodecki6}. However, the situation becomes more complex
when we distill the entangled state whose subsystems have higher
dimensions or that has a higher rank.

In this paper we show that the rank three bipartite entangled
states are distillable under LOCC. We give the concrete method of
distilling this class of states. It helps infer the analytical
calculation of distillable entanglement \cite{Rains,Devetak}. A
rank three state is entangled if and only if (iff) it is NPT,
namely there is no PPT BE state of rank three \cite{Lewenstein}.
So we also obtain that there are no rank-three NPT BE states and
all of them can be used for quantum-information tasks. It is
similar to the case of rank two states and we conclude: a rank two
or three state is distillable iff it is entangled. This conclusion
does not hold for the bipartite entangled states with higher
ranks, e.g., there have been the rank four PPT BE states
constructed by the unextendible product bases (UPB) \cite{Mor}.

Moreover, we will investigate the NPT states of rank four and find
out some families of states that are distillable. This helps
distill the NPT states which have more complex structure. In
addition, we will show that locally converting the Werner state
into the rank three entangled state is difficult, so our result is
independent of the expectant fact that there exists NPT BE Werner
state.

The rest of this paper is organized as follows. In Sec. II we
prove our main result on rank three states and then we use it to
distill the rank four NPT states. We also discuss the relationship
between the result in this paper and the Werner state. We conclude
in Sec. III.

\section{distillation of rank three and four bipartite states}

Throughout this paper we will use the following notations. The
rank of a bipartite state $\rho_{AB}$ is referred to as
$r(\rho_{AB})$, and the reduced density operator of it as
$\rho_A=\mbox{Tr}_B\rho_{AB},\rho_B=\mbox{Tr}_A\rho_{AB}$. The
range of the density operator $\rho_{AB}$ is referred to as
$R(\rho_{AB})$. Another useful tool is the so-called invertible
local operator (ILO) (or the local filter) \cite{Dur}, namely the
nonsingular matrix. Physically, it can be probabilistically
realized through the positive operator valued measure (POVM)
\cite{Nielsen}, so we can use it when distilling the NPT states.

We first consider the NPT states of rank three. Before proving our
main theorem, we recall a useful lemma that was proved in
\cite{Horodecki6}.

\textit{Lemma 1}. If
$r(\rho_{AB})<\mbox{max}[r(\rho_{A}),r(\rho_{B})]$, then the
bipartite state $\rho_{AB}$ is distillable.
\hspace*{\fill}$\blacksquare$

The lemma has been used to show that there is no rank two BE state
\cite{Horodecki6}. It was proven by using the reduction criterion
\cite{Horodecki5}, i.e., a state is distillable when the reduction
criterion is violated (See Eq. (6) in \cite{Horodecki6}). It
follows from lemma 1 that any rank three state in $M\times N$
space with $\mbox{max}[M,N]>3$ is distillable. Since an NPT state
in $2\times2$ or $2\times3$ space is also distillable
\cite{Bennett1,Bennett2,Horodecki2}, it suffices to consider the
rank three NPT states $\rho_{AB}$ in $3\times3$ space. Moreover,
we can perform some ILO on the subsystem $B$ such that
$\rho_B=\frac13I$. Then only the state having the following form
does not violate the reduction criterion ( up to local unitary
transformations )
\begin{equation}
\sigma_{AB}\equiv\frac13|\psi_0\rangle\langle\psi_0|+\frac13|\psi_1\rangle\langle\psi_1|
+\frac13|\psi_2\rangle\langle\psi_2|,\sigma_B=\frac13I
\end{equation}
where the three eigenvectors satisfy
$\langle\psi_i|\psi_j\rangle=\delta_{ij}$ and
\begin{eqnarray}
|\psi_0\rangle&=&\cos\theta|00\rangle+\sin\theta|11\rangle,\\
|\psi_1\rangle&=&\sum\nolimits^{2}_{i,j=0}b_{ij}|ij\rangle,\\
|\psi_2\rangle&=&\sum\nolimits^{2}_{i,j=0}c_{ij}|ij\rangle.
\end{eqnarray}
Notice that there is always at least a Schmidt rank two state by
linear combination of the eigenvectors. In addition, any spectral
decomposition of the state $\sigma_{AB}$ have the form in Eq. (1)
(in which the state $|\psi_0\rangle$ has a more general form,
e.g., $|\psi_0\rangle=\sum\nolimits^{2}_{i,j=0}a_{ij}|ij\rangle$).

In what follows we will concentrate on the NPT state $\sigma_{AB}$
in Eq. (1) because any rank three NPT state in $3\times3$ space
can be locally converted into $\sigma_{AB}$, otherwise it is
distillable in terms of the reduction criterion. There is a simple
situation we can treat easily as follows.

\textit{Lemma 2}. The state $\sigma_{AB}$ is distillable when
there is a product state in its range.

\textit{Proof.} Without loss of generality, we consider the state
$\sigma_{AB}$ with $\theta=0$. Then its coefficients
$b_{i0},c_{i0},i=0,1,2$ equal zero because of the condition
$\sigma_B=\frac13I$. We project the state $\sigma_{AB}$ by using
the local projector
$I_A\otimes(|1\rangle\langle1|+|2\rangle\langle2|)_B$ and obtain
the resulting state
$\frac12|\psi_1\rangle\langle\psi_1|+\frac12|\psi_2\rangle\langle\psi_2|$.
It's a rank two NPT state and hence distillable. It implies the
state $\sigma_{AB}$ is also distillable.
\hspace*{\fill}$\blacksquare$

Lemma 2 has given a criterion that tells whether a rank three NPT
state is distillable. We will generalize it to the case of rank
four states later. It is also useful for the distillation of
general rank three NPT state as shown below. Let us consider the
state $\sigma_{AB}$ whose range has no product state. We take the
projector $P_{AB}$ onto the $2\times3$ subspace spanned by
$\{|00\rangle,|01\rangle,|02\rangle,|10\rangle,|11\rangle,|12\rangle\}$
and obtain the state
\begin{equation}
\sigma^1_{AB}=|\psi^1_0\rangle\langle\psi^1_0|+|\psi^1_1\rangle\langle\psi^1_1|
+|\psi^1_2\rangle\langle\psi^1_2|,
\end{equation}
which is not normalized for convenience. The resulting states
$|\psi^1_i\rangle$ equal $P_{AB}|\psi_i\rangle$, respectively. We
will follow this notation below, e.g.,
$|\psi^2_i\rangle=V_A\otimes V_B|\psi^1_i\rangle$, etc.

The state $\sigma^1_{AB}$ is distillable if it is entangled since
it is in $2\times2$ or $2\times3$ space. Let us consider the case
in which $\sigma^1_{AB}$ is separable. First, the state
$\sigma^1_{AB}$ is in $2\times2$ space iff
$b_{i2}=c_{i2}=0,i=0,1$. In this case, the condition
$\sigma_B=\frac13I$ leads to
$b_{2i}b^*_{22}+c_{2i}c^*_{22}=0,i=0,1$ and
$|b_{22}|^2+|c_{22}|^2=1$. When $b_{22}c_{22}=0$, either the state
$|\psi_1\rangle$ or $|\psi_2\rangle$ becomes a product state and
hence $\sigma_{AB}$ is distillable in terms of lemma 2; When
$b_{22}c_{22}\neq0$, we can remove the coefficients
$b_{2i},c_{2i},i=0,1$ by using linear combination of the
eigenvectors $|\psi_i\rangle,i=0,1,2$. It is then easy to see that
$R(\sigma_{AB})$ contains a product state and thus $\sigma_{AB}$
is distillable.

Second, we investigate the state $\sigma^1_{AB}$ in $2\times3$
space. Notice the rank of $\sigma^1_{AB}$ remains three, otherwise
there will be a product state in $R(\sigma_{AB})$ and it is
distillable. We can always write a rank three separable state
$\rho$ in $2\times3$ space as the sum of three product states
\cite{Wootters,Werner}. To prove it, suppose the state has the
form
\begin{equation}
\rho=\sum\nolimits^{d-1}_{i=0}|\phi_i\rangle|\omega_i\rangle\langle\phi_i|\langle\omega_i|,d>3.
\end{equation}
Without loss of generality we choose the first three product
states as a set of linearly independent vectors, so any other
product state can be written as
$|\phi_j\rangle|\omega_j\rangle=\sum\nolimits^2_{i=0}k_{ij}|\phi_i\rangle|\omega_i\rangle,j=3,...$
Notice the vectors $|\omega_i\rangle,i=0,1,2$, and two vectors in
$|\phi_i\rangle,i=0,1,2$ are linearly independent, respectively.
So the product state $|\phi_j\rangle|\omega_j\rangle,j>3$ equals
either one of the first three product states, or
$|\phi_j\rangle|\omega_j\rangle=\sum\nolimits^1_{i=0}k_{ij}|\phi_i\rangle|\omega_i\rangle$
in which $|\phi_0\rangle$ is proportional to $|\phi_1\rangle$. In
this case it is easy to write the state $\rho$ as the sum of three
product states.

Using the above conclusion, we can express the state $\sigma_{AB}$
by means of eigenvectors
$|\psi_i\rangle=(a_{i0}|0\rangle+a_{i1}|1\rangle)|\phi_{i1}\rangle+|2\rangle|\phi_{i2}\rangle,i=0,1,2.$
Moreover, the vectors $|\phi_{i1}\rangle$'s are linearly
independent, while $|\phi_{i2}\rangle$'s linearly dependent. We
perform some ILO's on the state $\sigma_{AB}$ and remove two
coefficients $a_{00}$ and $a_{11}$. The resulting state
$\sigma^2_{AB}$ still has the form in Eq. (1), otherwise it is
distillable.

For the state $\sigma^2_{AB}$ when the condition $a_{20}a_{21}=0$
is satisfied, we find that $R(\sigma^2_{AB})$ contains a product
state because of the orthogonal conditions
$\langle\psi^2_i|\psi^2_j\rangle=\delta_{ij}$. So the state
$\sigma_{AB}$ is distillable. Let us move to investigate the state
$\sigma^2_{AB}$ satisfying the condition $a_{20}a_{21}\neq0$. By
performing ILO's on $\sigma^2_{AB}$ we greatly simplify its form
such that
\begin{equation}
\sigma^3_{AB}=|\psi^3_0\rangle\langle\psi^3_0|+|\psi^3_1\rangle\langle\psi^3_1|
+|\psi^3_2\rangle\langle\psi^3_2|,
\end{equation}
where
\begin{eqnarray}
|\psi^3_0\rangle&=&|00\rangle+|2\rangle|\psi\rangle,\\
|\psi^3_1\rangle&=&|11\rangle+|2\rangle|\phi\rangle,\\
|\psi^3_2\rangle&=&(|0\rangle+|1\rangle)|2\rangle+
|2\rangle(\alpha|\psi\rangle+\beta|\phi\rangle),\\
|\psi\rangle&=&x_0|0\rangle+x_1|1\rangle+x_2|2\rangle,\\
|\phi\rangle&=&y_0|0\rangle+y_1|1\rangle+y_2|2\rangle.
\end{eqnarray}
Notice the state is not normalized and the condition
$\sigma^3_B=\frac13I$ is also not required. We project
$\sigma^3_{AB}$ by the projector
$[|0\rangle(\langle0|+a\langle1|)+|2\rangle\langle2|]_A\otimes
I_B,a\in R$ and obtain the state $\sigma^4_{AB}$ in $2\times3$
space. It is entangled and thus distillable when its partial
transpose is not positive \cite{Peres}. Nevertheless, there may be
some cases in which the coefficients $x_i,y_i,i=0,1,2$ make that
$(\sigma^4_{AB})^{T_A}\geq0$. We are going to find out such
coefficients by calculating several average values
$\mbox{Tr}[(\sigma^4_{AB})^{T_A}|\omega_i\rangle\langle\omega_i|]$,
where
$|\omega_0\rangle=|00\rangle+b|22\rangle,|\omega_1\rangle=|00\rangle+b|21\rangle
,|\omega_2\rangle=|01\rangle+b|20\rangle,|\omega_3\rangle=|02\rangle+b|20\rangle,b\in
C.$ To keep the average value always positive, we find that it is
necessary that $x_1=x_2=0$. However, this means the state
$|\psi^3_0\rangle$ is of product form and hence the state
$\sigma^3_{AB}$ is distillable. As it can be converted into the
state $\sigma_{AB}$ by ILO's, the latter is also distillable. Now
we reach our main theorem in this paper.

\textit{Theorem.} The rank three NPT states are distillable under
LOCC. \hspace*{\fill}$\blacksquare$

So the rank three entangled states can be used for
quantum-information tasks. In fact, we have proposed the method of
distilling $\sigma_{AB}$ in the proof of the theorem. First, when
the given state contains a product state in its range, it can be
projected onto a rank two entangled state. According to the
reduction criterion, we can distill it by the procedure similar to
the famous BBPSSW protocol \cite{Bennett2,Horodecki5}. It is also
the method of distilling the rank three entangled states that
cannot be converted into $\sigma_{AB}$. Second, when the given
state $\rho$ contains no product state in $R(\rho)$, we project it
by the projector $(|0\rangle\langle0|+|1\rangle\langle1|)_A\otimes
I_B$. The resulting state is entangled and thus distillable;
otherwise, we should project the initial state $\rho$ by the
projector
$[|0\rangle(\langle0|+a\langle1|)+|2\rangle\langle2|]_A\otimes
I_B$ after performing some ILOs on $\rho$. There will be a
suitable parameter $a$ making the resulting state entangled and
thus distillable.

The rank three entangled states are a quite special class of
states. As there have been PPT BE states of any higher rank (e.g.,
rank four PPT BE states constructed by UPB \cite{Mor}), we indeed
have found out the lowest rank space in which there is no BE
state. It also implies that when a state can be locally projected
into some rank three NPT state, then it is distillable. This
causes new methods of distilling quantum states having more
complex structure. We will show it in terms of distilling the rank
four NPT states below. One may also find other way to distill the
entangled states based on the theorem. For example, the tensor
product of the rank three entangled states are also entangled for
certain.

On the other hand, the analytical calculation of distillable
entanglement is also an important issue in quantum information
theory. The problem is very difficult and there have been some
optimal bounds on distillable entanglement \cite{Rains,Devetak}.
Specially, the bound is saturated if we can find a way to distill
the state and get the same amount of pure entanglement as the
bound. In this case we get the analytical result of distillable
entanglement. As it is possible to find out whether the bound on
rank three NPT state is saturated by using our method of
distilling it, we indeed provide new ways to calculate the
distillable entanglement.

Third, our result is also independent of the expectant fact that
there exist NPT Werner states $\rho_w$. We do not know whether an
NPT state $\rho$ is distillable, even it can be converted into
some Werner state which is proved to be not distillable under
LOCC. One can easily exemplify it by locally taking some rank
three NPT state into $\rho_w$, while the latter is expected to be
not distillable. Conversely, it is difficult to convert the Werner
state into the state $\sigma_{AB}$, so we still do not know
whether the latter is distillable. To see it, we have the Werner
state in a $N\times N$ space as follows \cite{Werner}
\begin{eqnarray}
\rho_w&=&(a+b)\sum\nolimits^{N-1}_{i,j=0}|ij\rangle\langle
ij|\nonumber\\
&-&2b\sum\nolimits^{N-1}_{i<j=0}\frac{|ij\rangle-|ji\rangle}{\sqrt2}\frac{\langle
ij|-\langle ji|}{\sqrt2},
\end{eqnarray}
where $a>0,b<0$ are two parameters satisfying $a+b\geq0$. The most
general local transformation on a quantum state $\rho$ has the
form $\Lambda(\rho)=\sum\nolimits_i A_i\otimes B_i\rho
A^{\dag}_i\otimes B^{\dag}_i$ \cite{Vedral}. Because the resulting
state is entangled, there must be at least one pair of Kraus
operators $A_i,B_i$ that have at least rank two, respectively. In
this case, the state $\Lambda(\rho)$ will have the rank not less
than four when $a+b>0$, which means a rank three NPT state cannot
be output by this local channel. The only exception happens when
$a+b=0$, but it is difficult to judge whether the state
$\Lambda(\rho)$ is of rank three and entangled.

Let us investigate further the problem of distilling rank four
states by using the theorem in this paper. Different from the case
of rank three state, it is well-known that there indeed exist PPT
BE states of rank four even in the $3\times3$ space. It is easy to
show that the NPT BE states of rank four possibly exist only in
three kinds of spaces, $4\times4,3\times4,3\times3$ in terms of
lemma 1. One will meet lots of difficulties when applying the
technique in this paper to distill the rank four NPT state $\rho$,
e.g., the resulting state from $\rho$ by projection can be
$3\times3$ and it may be PPT BE. Besides, the Peres-Horodecki
criterion is no more a sufficient condition for the separability
of state in $2\times4$ space, etc. Nevertheless, we still can
obtain some useful results on this problem when the target state
has a special form.

\textit{Lemma 3}. For a rank four NPT state in $4\times4$ or
$3\times4$ space, it is distillable when there is a product state
in its range.

\textit{Proof.} By employing similar deduction for the state
$\sigma_{AB}$, only the state having the following form does not
violate the reduction criterion
\begin{equation}
\rho_{AB}=\frac14\sum\nolimits^3_{i=0}|\psi_i\rangle\langle\psi_i|,\rho_B=\frac14I,
\end{equation}
where the four eigenvectors satisfy
$\langle\psi_i|\psi_j\rangle=\delta_{ij}$. Up to the local unitary
transformations we have $|\psi_0\rangle=|00\rangle$. Next, we
project the state $\rho_{AB}$ by the projector
$I_A\otimes(|1\rangle\langle1|+|2\rangle\langle2|+|3\rangle\langle3|)_B$
and obtain the NPT state
$\rho^{\prime}_{AB}=\frac13\sum\nolimits^3_{i=1}|\psi_i\rangle\langle\psi_i|$
in $2\times3$, or $3\times3$, or $4\times3$ space. By means of the
BBPSSW and Horodeckis' protocol, our theorem and the reduction
criterion, respectively, the state $\rho^{\prime}_{AB}$ and hence
$\rho_{AB}$ is always distillable. \hspace*{\fill}$\blacksquare$

So we have generalized lemma 1 to the case of rank four NPT
states. Moreover, we hope that it always holds for the NPT states
whose rank equal to its maximal dimension of subsystems. However,
it does not hold when the rank of a state is larger, e.g, the PPT
BE state in $3\times3$ space constructed in \cite{Horodecki4}
contains infinitely many product states in its range, but its rank
equals eight. It is also unclear that whether the rank four NPT
states $\rho$ in this space are distillable. Solving this problem
is more difficult since we cannot rely on the reduction criterion.
However, $\rho$ is distillable when we can project it onto a rank
three NPT state in terms of our theorem.

For example, the following $3\times3$ rank four NPT state is
distillable
\begin{eqnarray}
\rho_{AB}&=&\lambda_0|00\rangle\langle00|+\lambda_1|01\rangle\langle01|+
\lambda_2|\psi_2\rangle\langle\psi_2|+\lambda_3|\psi_3\rangle\langle\psi_3|,\nonumber\\
|\psi_2\rangle&=&\sum\nolimits^{2}_{i,j=0}c_{ij}|ij\rangle,\nonumber\\
|\psi_3\rangle&=&\sum\nolimits^{2}_{i,j=0}d_{ij}|ij\rangle,\lambda_0,\lambda_1,\lambda_2,\lambda_3>0.
\end{eqnarray}
To prove it, we project the state $\rho_{AB}$ by the projector
$(|1\rangle\langle1|+|2\rangle\langle2|)_A\otimes I_B$. When the
resulting state
$\rho^{1}_{AB}=\lambda_2|\psi^1_2\rangle\langle\psi^1_2|+\lambda_3|\psi^1_3\rangle\langle\psi^1_3|$
is entangled, it is also distillable. On the other hand when
$\rho^{1}_{AB}$ is separable, we can write it as the sum of two
product states since it is in a space not larger than $2\times3.$
Besides, the rank of $\rho^{1}_{AB}$ must be two because of
$r(\rho_A)=3$. By performing some ILOs on the state $\rho_{AB}$
and linear combination of $|\psi_2\rangle$ and $|\psi_3\rangle$,
we can convert them into
$|\psi^2_2\rangle=|0\rangle|\phi_0\rangle+|1\rangle|\phi_1\rangle$
and
$|\psi^2_3\rangle=|0\rangle|\omega_0\rangle+|2\rangle|\omega_2\rangle$,
and keep the other two terms $|00\rangle$ and $|01\rangle$
unchanged.

When either of the states $|\psi^2_2\rangle$ and
$|\psi^2_3\rangle$ is of product form, we easily project the state
$\rho^2_{AB}$ onto a $2\times3$ subspace, the resulting state is
still entangled and distillable. On the other hand when both the
states $|\psi^2_2\rangle$ and $|\psi^2_3\rangle$ are entangled, we
project the state $\rho^2_{AB}$ by the projector
$[|0\rangle(\langle0|+a\langle1|)+|2\rangle\langle2|]_A\otimes
I_B$. The obtained state $\rho^3_{AB}$ is $2\times3$ and its rank
is four by choosing suitable parameter $a$. This state is
separable iff it has the decomposition
$\rho^3_{AB}=|\psi\rangle_A\langle\psi|\otimes|\omega_2\rangle_B\langle\omega_2|
+|0\rangle_A\langle0|\otimes\rho^3_B$ with $r(\rho^3_B)=3.$
However it is impossible, since it requires a $4\times4$
coefficient unitary matrix $[a_{ij}]$ in which $a_{i3}=0,i=1,2,3,$
and $a_{0i},i=0,1,2$ cannot be zero simultaneously. Hence the
state $\rho^3_{AB}$ is entangled and thus distillable. This also
completes the proof showing that the state $\rho_{AB}$ in Eq. (15)
is distillable.

As above we have given several families of states that can be
distilled by means of the fact that the rank three NPT states are
distillable. The main difficulty in entanglement distillation is
the great amount of parameters that cannot be removed during the
filtering process. For example, it is unknown that whether the
rank four NPT states are distillable. All in all, more efforts are
required to distill other classes of rank four NPT states.

\section{conclusions}

We have proved that the bipartite rank three NPT states and some
families of rank four NPT states are distillable. So they are
indeed available resource for quantum-information tasks. An open
problem is that whether all rank four NPT states are distillable.
Our result also gives an insight into the relationship between the
low rank states and the Werner states.

The work was partly supported by the NNSF of China Grant
No.90503009, No.10775116, and 973 Program Grant No.2005CB724508.

\end{document}